\documentclass[12pt]{article}

\newcommand{\bra}[1]{\langle{#1}|}
\newcommand{\ket}[1]{|{#1}\rangle}

\def\beq{\begin{equation}}
\def\eeq{\end{equation}}
\def\beqa{\begin{eqnarray}}
\def\eeqa{\end{eqnarray}}

\newcommand{\eq}[1]{Eq. (\ref{#1})}

\newcommand{\e}{\epsilon}
\newcommand{\as}{\alpha_s}
\renewcommand{\b}{\beta}

\def\ifm{\ifmmode}

\def\M{{\cal M}}

\begin{document}

\begin{titlepage}

\rightline{DFTT-16/2008}

\rightline{\hfill June 2008}

\vspace{3.5cm}

\centerline{\Large \bf All-order results for soft and collinear gluons} 

\vskip 1.2cm

\centerline{\bf Lorenzo Magnea\footnote{e-mail: {\tt magnea@to.infn.it}}}
\centerline{\sl Dipartimento di Fisica Teorica, Universit\`a di Torino}
\centerline{\sl and INFN, Sezione di Torino}
\centerline{\sl Via P. Giuria 1, I--10125 Torino, Italy}

\vskip 2.5cm
 
\begin{abstract}

\noindent I briefly review some general features and some recent 
developments concerning the resummation of long-distance singularities 
in QCD and in more general non-abelian gauge theories.  I emphasize the 
field-theoretical tools of the trade, and focus mostly on the exponentiation 
of infrared and collinear divergences in amplitudes, which underlies the 
resummation of large logarithms in the corresponding cross sections. I then
describe some recent results concerning the conformal limit, notably 
the case of ${\cal N} = 4$ superymmetric Yang-Mills theory.

\end{abstract}

\end{titlepage}

\newpage

\section{Introduction}

Massless gauge field theories, which are classically conformal invariant, 
are characterized by the fact that all length scales enter in the calculation
of any given physical process. Consider for example the perturbative 
calculation of a scattering amplitude ${\cal A} (p_1, \ldots , p_n)$ with
momentum invariants characterized by a common scale $Q$. Just like 
in any quantum theory, when we compute ${\cal A}$ beyond the leading
perturbative order, we must allow for exchanges of virtual quanta of 
arbitrarily high energy, $E \gg Q$, corresponding to processes happening at arbitrarily small length scales. These exchanges are responsible for the ultraviolet
(UV) problems we encounter in perturbative calculations, and must be dealt 
with, when possible, with renormalization. If our theory is massless (or if the 
masses are negligible with respect to the scale $Q$), a similar problem arises 
at the other end of the spectrum: we must allow for exchanges of 
very low-energy quanta, $E \ll Q$, which happen at very
large distances. Such processes may or may not, in a general field theory,
endanger our calculational framework, but they certainly do so in the case 
of gauge theories, even abelian ones. This is the origin of the infrared (IR) 
problems of perturbative calculations, which are usually dealt with using factorization. 

In either case, it is worth recalling that the appearance of infinities in our perturbative calculations is due to the fact that we have stretched our approximations beyond their limits of applicability. In the UV regime, we 
have tacitly assumed that our theory should be applicable at extreme 
short distances, where in fact we do not even know what the relevant 
degrees of freedom might be. In the case of renormalizable theories, we 
are forgiven our arrogance, since we can show that physics at very short 
distances effectively decouples from the calculation of our amplitude at 
the scale $Q$: the contributions of high-energy quanta factorize, and can 
be absorbed into rescalings of the local couplings of our original theory.
In the IR regime, for gauge theories, our mistake is more subtle: massless
gauge bosons mediate long-range interactions, which cannot be switched 
off even at asymptotycally large distances; hence, it is not correct to 
formulate our perturbative expansion in a Hilbert space of Fock states 
built with creation and annihilation operators of free fields. The true 
asymptotic states of the theory are coherent states containing an infinite 
number of massless quanta, and the price of stretching our approximation 
is that the $S$ matrix actually does not exist in our Hilbert space: all 
scattering amplitudes diverge because of virtual IR exchanges.

In QED, and to a more limited extent in QCD, IR problems are alleviated
by the KLN theorem~\cite{Kinoshita:1962ur,Lee:1964is}. When we 
compensate for our inadequate choice of Hilbert space by constructing 
physically measurable probabilities, which are obtained by summing over
all Fock states that are degenerate in energy, all IR divergences must 
cancel. This is sufficient to solve the IR problem in QED, where actually
a sum over final-state degeneracies is enough to achieve the required 
cancellation. In an unbroken non-abelian gauge theory like QCD, things 
are considerably more complicated: because of confinement, the relationship
between partonic Fock states and the true non-perturbative asymptotic 
states of the theory (which are color-singlet hadrons) is highly non-trivial,
and beyond the range of our techniques; this is reflected at the perturbative
level in the fact that a sum over initial state degeneracies is both necessary
to cancel divergences, and impossible to perform in practice, since initial state partons are far from free at large distances.

In order to rescue the applicability of perturbative methods, one must resort 
once again to factorization, coupled with asymptotic freedom. One exploits
the quantum-mechanical incoherence of processes happening at different 
distance scales to show that high-energy inclusive cross sections can be 
written as convolutions of short-distance finite partonic cross sections, which 
can be computed in perturbation theory thanks to asymptotic freedom, with 
long-distance factors (parton distributions or fragmentation functions), which 
are non-perturbative but universally associated with hadronic wave functions.

Proofs of factorization are highly non-trivial in perturbation 
theory~\cite{Collins:1989gx}, but they pay big dividends. First of all, they
underpin essentially all perturbative QCD predictions for high-energy cross 
sections, from Deep Inelastic Scattering, to Drell-Yan production of 
electroweak vector bosons and Higgs bosons, to general jet cross sections.
There are, furthermore, several other applications, important both for theory 
and for phenomenology, some of which will be reviewed below.

\begin{itemize}

\item Factorization leads to evolution equations. All factorization theorems
introduce intermediate arbitrary scales separating the momentum space
regions one wishes to disentangle. Demanding that physical observables be independent of these arbitrary scales leads to evolution equations for individual
contributions to the factorized observable. A prime example is of course
Altarelli-Parisi~\cite{Altarelli:1977zs} evolution of parton distributions.

\item Solving the evolution equations dictated by factorization leads to the resummation of classes of logarithmic contributions to all orders in perturbation theory. Renormalization group evolution and collinear evolution of parton
distributions are examples, but also Sudakov resummation, both for threshold
and transverse momentum logarithms, can be derived in this way 
(see~\cite{Contopanagos:1996nh} and, for a review, \cite{Laenen:2004pm}).
Such resummations are important, in some cases essential, for the
phenomenological success of perturbative QCD. Alternatively, having 
an established a factorization theorem, one can apply effective field theory
methods to derive the same results in a different systematical way (see,
for example,~\cite{Becher:2006nr}).

\item Resummation, in turn, probes the all-order structure of perturbation 
theory and helps identify leading non-perturbative corrections to many cross
sections. The study of power-suppressed corrections to QCD factorization 
theorems has in fact developed into an active subfield of research, with many phenomenological applications and new interesting theoretical ideas. The 
literature is vast, involving renormalon techniques as well as resummations;
for reviews and further references on some of these ideas see for 
example~\cite{Dasgupta:2003iq} and~\cite{Gardi:2006jc}.

\item At the amplitude level, resummation of IR singularities displays
universal features of gauge theories which find application both in theory 
and in phenomenology. Understanding the structure of infrared and collinear 
poles at high orders is instrumental to construct subtraction algorithms to
compute efficiently multiparticle cross sections at colliders. On the other 
hand, remarkably, the all-order structure of infrared poles, uncovered in QCD,
has recently found application in the context of supersymmetric gauge 
theories, and most notably for ${\cal N} = 4$ Super-Yang-Mills (SYM) 
theory, which is conjectured to be equivalent to string theory on the
background of an ${\rm AdS}_5 \times S^5$ space-time. Gluon amplitudes
can now be computed in weak-coupling ${\cal N} = 4$ SYM to very high
order, and can also, in some cases, be computed at strong ('t Hooft) coupling
using directly the AdS-CFT correspondence~\cite{Alday:2007hr}. The universal
structure of infared and collinear singularities of these amplitudes places 
powerful constraints on their structure, and helps identify the relevant 
anomalous dimensions. This is a very active and fast-developing field, 
recently reviewed in~\cite{Dixon:2008tu} and~\cite{Alday:2008cg}.

\end{itemize} 

In the following sections, I will briefly describe some of the field theory 
tools that are used to study the universal structure of gauge theories at
long distances. I will begin in Section 2 by discussing qualitatively the 
connection between factorization, evolution ad resummation; then,
in Section 3, I will introduce some of the technical tools that are needed
to make precise all-order statements. In Section 4, I will focus on the 
simplest example of nonabelian exponentiation of infrared singularities, the
form factor of a massless particle in a gauge theory. Finally, in Section 5,
I will describe some applications of these results to conformal gauge 
theories. I will include in the discussion several new results recently 
obtained in~\cite{Dixon:2008gr}.

\section{Factorization leads to resummation}
\label{facto}

A factorization theorem is a sufficient (though not necessary) condition
to perform a resummation of perturbation theory, typically through the solution
of an appropriate evolution equation. Clearly, the difficult work is proving factorization to all orders in perturbation theory, which requires a detailed
diagrammatic analysis and often a delicate implementation of the symmetry properties of the theory. Once factorization is established, evolution equations
automatically follow, and their solutions entail an all-order organization of certain perturbative contributions. Let's look at three classical examples,
following~\cite{Contopanagos:1996nh}.

\begin{itemize}

\item The prototypical perturbative factorization is the renormalization
of UV divergences. In this case the difficult step is to prove, to all orders, that
the dependence of a generic Green function of the theory on the cutoff scale
can be associated with a finite number of universal multiplicative factors. If that's the case, then one may write, for an $n$-point bare correlator $G_0^{(n)}$,
\beq
G_0^{(n)} \left(p_i, \Lambda, g_0 \right) = \prod_{i = 1}^n Z_i^{1/2}
\left(\Lambda/\mu, g(\mu)\right) ~G_R^{(n)} \left(p_i, \mu, g (\mu)
\right)~,
\label{rinorm}
\eeq
where $g_0$ is the bare coupling, and $\Lambda$ the UV cutoff, which can 
be interpreted as the scale of `new physics', or the scale at which the 
effective low-energy theory under consideration breaks down. In order to 
perform the factorization in \eq{rinorm} it has been necessary to introduce
and intermediate, arbitrary scale $\mu$. Bare Green functions, however, do 
not depend on $\mu$, so that we can write
\beq
\frac{d G_0^{(n)}}{d \mu} = 0 ~~ \rightarrow ~~
\frac{d \log G_R^{(n)}}{d \log \mu} = - \sum_{i = 1}^n \gamma_i
\left( g(\mu) \right)~,
\label{rinorm2}
\eeq
where $\gamma_i (g) \equiv d \log Z_i/ d \log \mu^2$. Solving the renormalization group equation, \eq{rinorm2}, resums the logarithmic 
dependence on the renormalization scale $\mu$. Furthermore, since
renormalized Green functions, up to their overall engineering dimension,
depend on $\mu$ only through ratios such as $p_i \cdot p_j/\mu^2$,
the same equation can be used to extract information about the dependence
of $G_R^{(n)}$ on external momenta. Notice that, in order to enforce the cancellation in \eq{rinorm2}, the anomalous dimensions $\gamma_i (g)$ 
can only depend on arguments that are common to $Z_i$ and  to 
$G_R^{(n)}$, the only one in this case being the renormalized coupling 
$g$.

\item Another familiar example is collinear factorization in high-energy QCD
cross sections, most topically DIS. In this case the difficult task is to show 
that collinear singularities in the cross section can be absorbed into universal factors associated with the wave functions of initial state hadrons. For DIS structure functions, say $F_2 (x, Q^2)$, this is true in the form of a convolution,
which becomes an ordinary product upon taking a Mellin transform. One writes 
then
\beq
\widetilde{F}_2 \left(N, \frac{Q^2}{m^2}, \alpha_s \right) = 
\widetilde{C} \left(N, \frac{Q^2}{\mu_F^2}, \alpha_s \right) 
\widetilde{f} \left(N, \frac{\mu_F^2}{m^2}, \alpha_s \right)~.
\label{disfa}
\eeq
Here $m$ is a label for a collinear regulator, say a light quark mass;
$\widetilde{C}$ is a perturbatively computable coefficient function,
free of collinear sensitivity, while $\widetilde{f}$ is a universal (but non-computable) parton distribution. Again, to perform factorization it has ben necessary to introduce an arbitrary scale $\mu_F$, and one can exploit
the fact that the structure function $\widetilde{F}_2$ does not depend in the choice of $\mu_F$. One derives
\beq
\frac{d \widetilde{F}_2}{d \mu_F} = 0 ~~ \rightarrow ~~
\frac{d \log \widetilde{f}}{d \log \mu_F} = \gamma_N \left( \alpha_s \right)~,
\label{disfa2}
\eeq
where $\gamma_N \left( \alpha_s \right) \equiv - d \log \widetilde{C}/d 
\log \mu_F$ are the Mellin moments of the appropriate Altarelli-Parisi 
splitting function. The anomalous dimension $\gamma_N$ can only depend
on arguments common to $\widetilde{C}$ and $\widetilde{f}$, in this case 
$N$ and $\as$. Solving \eq{disfa2} resums logarithms of the factorization 
scale, and allows to evolve parton distributions to the scales appropriate
for applications to other high-energy cross sections.

\item Let us finally turn to the most significant (and difficult) case of 
Sudakov resummation. In the previous two cases one was dealing with the resummation of single logarithms, arising from single (UV or collinear) poles
of the corresponding amplitudes. Sudakov resummation involves double (IR and collinear) poles, which requires in principle a more elaborate factorization (see,
however, the arguments in~\cite{Forte:2002ni}). Since our focus below will be 
on amplitudes rather than cross sections, let us begin by considering the 
simplest scattering amplitude which is affected by such double poles, the 
form factor of any massless particle minimally coupled to a massless gauge 
boson. Using a massless quark as an example one can define
\beq
\Gamma_\mu (p_1, p_2; \mu^2, \e) \equiv \bra{{p_1,p_2}}
J_\mu (0) \ket{0} =  \overline{u}(p_1) \gamma_\mu
v(p_2) ~\Gamma \left(\frac{Q^2}{\mu^2}, \as (\mu^2), \epsilon \right)~,
\label{fofadef}
\eeq
corresponding to pair creation of a $q \bar{q}$ pair out of the QCD vacuum 
by means of a source (an off-shell photon) of mass $Q$. It can be shown
(as reviewed in~\cite{Collins:1989bt}) that this amplitude factorizes into the product of different functions, each one responsible for a specific set of 
singularities. In dimensional regularization, the precise form of this factorization
can be written as~\cite{Dixon:2008gr}
\beqa
  \Gamma \left( \frac{Q^2}{\mu^2}, \as(\mu^2), \e \right) & = &
  C \left( \frac{Q^2}{\mu^2}, \frac{(p_i \cdot n_i)^2}{n_i^2 \mu^2}, 
  \as (\mu^2), \e \right) \, \times \,
  {\cal S} \left( \beta_1\cdot \beta_2, \as (\mu^2), \e \right)
  \nonumber \\ && \times \,
  \prod_{i = 1}^2 \left[ 
  \frac{J \left( \frac{(p_i \cdot n_i)^2}{n_i^2 \mu^2},\as (\mu^2), \e \right)}
  {{\cal J} \left( \frac{(\beta_i \cdot n_i)^2}{n_i^2}, \as(\mu^2), \e \right)} 
  \right]~.
\label{factorize}
\eeqa
Here $\beta_i$ are four velocities associated with the quark and the antiquark
(so that $p_i^\nu \propto Q \beta_i^\nu$), while $n_i$ are auxiliary vectors
associated with Wilson lines (to be described below), which are introduced in 
order to factorize wide-angle soft radiation from collinear one. The function
${\cal S}$ is an eikonal function responsible for the radiation of soft gluons, 
while the jet function $J$ is associated with radiation collinear to either the 
quark or the antiquark, and $C$ is finite as $\e \to 0$. Including both 
${\cal S}$ and $J$ double counts the soft-collinear region, which is 
compensated for by introducing eikonal versions of the jets, ${\cal J}$. In 
order to derive evolution equations, one can now exploit the renormalization 
group invariance of $\Gamma$, which must not depend on the scale $\mu$, 
as well as the manifest independence on the auxiliary vectors 
$n_i$. Depending on how the calculation is performed, the independence of 
$\Gamma$ on $n_i$ can be understood either as gauge invariance (if working
in an axial gauge), or as Lorentz invariance (if working in Feynman gauge, as
we will do below).

\end{itemize}

In order to move forward and be more precise we need at this point to step
back and introduce some technical tools. Specifically, in order to give precise operator expressions for the functions entering \eq{factorize}, we will need 
to introduce Wilson line operators, which are also instrumental in the mapping
to strong coupling which becomes possible for ${\cal N} = 4$ SYM. Furthermore,
in order to perform a consistent resummation in dimensional regularization, we
need to define the running coupling in $d = 4 - 2 \e$ dimensions. We briefly
turn to these issues in the next section.

\section{Tools of the trade}
\label{tools}

\subsection{Dimensional regularization for the strong coupling} 
\label{dimreg}

Dimensional regularization (DR), in its various flavors, is a unique tool for the 
study of non-abelian gauge theories. It can be used both for UV and IR
divergences, it preserves gauge invariance, it is by far the simplest scheme
to use from the computational point of view. In the context of all-order
calculations, it has further virtues. In this case, one starts with the 
renormalized theory, and regulates long-distance singularities by taking
$d = 4 - 2 \e$, with $\e < 0$. One must then recall that RG equations
acquire $\e$ dependence in $d \neq 4$. The coupling, for example, runs
according to
\beq
 \mu \, \frac{\partial \overline{\alpha}}{\partial \mu} \, \equiv \,
 \b (\e, \overline{\alpha}) \, = \, - \, 2 \e 
 \overline{\alpha} \, + \, \hat{\b} (\overline{\alpha}) ~,~~~ 
 \hat{\b} (\overline{\alpha}) \, = \, - \, \frac{\overline{\alpha}^2}{2
 \pi} \, \sum_{n = 0}^\infty \, b_n \left(
 \frac{\overline{\alpha}}{\pi} \right)^n \, . 
\label{betadim}
\eeq
The $\e$ dependence of the $\beta$ function is a consequence of the 
engineering dimension of the bare coupling, and it implies that the running 
coupling behaves like a power of its scale, $\as(\mu^2)/ \as(\mu_0^2) 
\sim (\mu^2/\mu_0^2)^{- \e}$: in fact, in $d >4$, the $\beta$ function
has an IR-free fixed point at $\as = 0$, where it vanishes with a positive 
derivative. As a consequence, $\as (\mu^2 = 0) = 0$. The RG equation, 
\eq{betadim}, is easily solved at one loop, yielding
\beq
 \overline{\alpha} \left(\mu^2, \e \right) =
 \alpha_s (\mu_0^2) \left[ \left(\frac{\mu^2}{\mu_0^2} \right)^\epsilon -
 \frac{1}{\epsilon} \left( 1 - \left(\frac{\mu^2}{\mu_0^2}
 \right)^\epsilon \right) \frac{b_0}{4\pi} \, \alpha_s(\mu_0^2)
 \right]^{-1}~.
 \label{olas}
\eeq
Note that $\overline{\alpha}$ depends only on the scale $\mu^2$ and on 
$\e$, but not on the chosen initial condition $\mu_0^2$. At higher orders an
explicit analytic solution such as \eq{olas} is not available, but one may still
expand $\overline{\alpha}$ in powers of the coupling at a fixed reference 
scale, as
\beqa
\hspace{-1mm}  \overline{\alpha} \left( \xi^2, \e \right) & = &
  \alpha_s \, \xi^{- 2 \e} +
  \alpha_s^2  \, \xi^{- 4 \e} \, \frac{b_0}{4 \pi \e}
  \left( 1 - \xi^{2 \e} \right)  \nonumber \\
  &&  \hspace{-8mm} + \,\, \alpha_s^3 \, \xi^{- 6 \e} \,
  \frac{1}{8 \pi^2 \e} \left[ \frac{b_0^2}{2 \e}
  \left( 1 - \xi^{2 \e} \right)^2 +
  b_1 \left( 1 - \xi^{4 \e} \right)
  \right] + {\cal O} \left( \as^4 \right) ~. 
 \label{highoas} 
 \eeqa
The key advantage of ~\eq{betadim} is that it provides a simple initial condition
for the solution of evolution equations for amplitudes, basically stating that
all radiative corrections vanish when the scale vanishes. This fact was first 
exploited to give an explicit exponentiated expression for the Sudakov form 
factor in~\cite{Magnea:1990zb}. A further advantage of~\eq{betadim} is 
that the Landau pole for the running coupling acquires a non-vanishing imaginary
part when $\e < - b_0 \as(\mu_0^2)/(4 \pi)$, a fact that can be exploited to
evaluate resummed amplitudes explicitly as analytic functions of the coupling 
and of $\e$~\cite{Magnea:2000ss}.

\subsection{Wilson lines and the eikonal approximation}
\label{WIlson}

An important feature of \eq{factorize} is the fact that all singular factors 
comprising the form factor have well-defined operator expressions. This 
is especially significant when one is trying to make a connection to 
non-perturbative features of the theory, as is the case for ${\cal N} = 4$
SYM. It is well-known, and easily verified, that in the soft approximation, 
relevant for the calculation of infrared poles, gluon interactions with other
hard partons can be completely expressed in terms of correlators of Wilson 
lines: energetic partons do not recoil against soft radiation, so that the only 
effect of interactions with soft gluons is the buildup of an eikonal phase
on the parton field; soft gluons, in turn, are only sensitive to the direction 
and color representation of the hard parton, but not to its spin and energy.
The situation with collinear gluons is not as simple:  it can be shown that they couple eikonally to hard partons moving in different light-cone directions, but 
they retain to some extent the spin and energy dependence of the coupling to partons belonging to their own jet. In either case, eikonal lines play a major 
role in factorization formulas such as \eq{factorize}. Defining the Wilson line
operator as
\beq
  \Phi_n (\lambda_2, \lambda_1) =
  P \exp \left[\, {\rm i} g \int_{\lambda_1}^{\lambda_2} d \lambda \,
  n \cdot A(\lambda n) \, \right]~,
\eeq
we can give explicit operator expressions for all the functions appearing in 
\eq{factorize}. The soft function ${\cal S}$ is just the eikonal approximation 
of the full form factor
\beq
  {\cal S} \left( \beta_1 \cdot \beta_2, \as (\mu^2), \e \right) =
  \langle 0 | \Phi_{\beta_2} (\infty,0) \, \Phi_{\beta_1} (0, - \infty) 
  \, | 0 \rangle~;
\label{calSdef}
\eeq
the jet functions $J$, on the other hand, couple a hard parton to an eikonal 
line off the light cone, along an arbitrary space-like direction $n^\mu$,
\beq
  J \left( \frac{(p \cdot n)^2}{n^2 \mu^2}, \as (\mu^2), \e \right) \, 
  u (p) \, = \, \langle 0 \, | \Phi_n (\infty, 0) \, \psi (0) \,  | p \rangle\, .
\label{Jdef}
\eeq
Eikonal jets ${\cal J}$, finally, represent the soft approximation of the 
partonic jets $J$, so that the parton field $\psi$ is replaced by its own Wilson 
line,
\beq
  {\cal J} \left( \frac{(\beta \cdot n)^2}{n^2}, \as(\mu^2), \e 
  \right) = \langle 0 | \Phi_n (\infty, 0) \, \Phi_\beta (0, - \infty)
  \, | 0 \rangle~.
\label{calJdef}
\eeq
As we will briefly summarize below, while the evolution 
equations~(\ref{rinorm2}) and~(\ref{disfa2}) were stemming 
from the invariance of the observable before factorization with 
respect to variations of a mass scale, in the case of \eq{factorize} one 
derives evolution by demanding invariance with respect to the choice of 
the `factorization vectors' $n_i^\mu$.

\section{Resummation for the form factor}
\label{fofa}

The derivation of the evolution equation for the form factor can be understood 
by considering $n_i^\mu$ dependence in \eq{factorize}. Clearly, all such dependence is through the dimensionless ratio $x_i \equiv (\beta_i 
\cdot n_i)^2/n_i^2$. Demanding that $\partial \log \Gamma/\partial 
\log x_i = 0$, and noting that $\Gamma$ depends on $x_i$ only through
the jet functions and the finite coefficient function $C$, one derives
\beqa
  x_i \, \frac{\partial}{\partial x_i} \log J_i & = &
  \ - \, x_i \, \frac{\partial}{\partial x_i} \log C  \, + \,
  x_i \, \frac{\partial}{\partial x_i} \log {\cal J}_i \nonumber \\
  & \equiv & \frac{1}{2} \left[{\cal G}_i \Big( x_i, \as(\mu^2), \e \Big)
  + {\cal K} \Big( \as(\mu^2), \e \Big)  \right] \, ,
\label{dJdn}
\eeqa
where the second line defines the functions ${\cal G}_i$ and ${\cal K}_i$.
The key feature of \eq{dJdn} is that the $n_i^\mu$ dependence of the partonic 
jets has been organized in a function ${\cal G}_i$, which carries all the 
kinematic dependence, but is finite as $\e \to 0$ (because $C$ is finite), 
plus a function ${\cal K}_i$, which on the contrary is a pure counterterm 
(because ${\cal J}$ is), but carries no kinematic dependence. Note also
that, while ${\cal J}_i$ has a double pole, its derivative with respect to $x_i$
must have only a single pole, since the double pole is independent of 
kinematics. Integrating \eq{dJdn} thus leads to one of the key features of resummation: double pole observables exponentiate, and their logarithms
contain only single poles.

It is not difficult to generalize the argument leading to \eq{dJdn} to the full 
form factor~\cite{Collins:1989bt}. One finds an equation of the same form
\beq
  Q^2 \frac{\partial}{\partial Q^2} 
  \log \left[\Gamma \left( \frac{Q^2}{\mu^2}, 
  \as(\mu^2), \e \right) \right] = \frac{1}{2} \left[ K \left(\e, \as(\mu^2) 
  \right) + G \left(\frac{Q^2}{\mu^2}, \as(\mu^2), \e \right) \right]  ,
\label{fofaeveq}
\eeq
where $G$ is finite as $\e \to 0$ and carries the full $Q^2$ dependence,
while $K$ is a $Q^2$-independent pure counterterm. In order to solve 
\eq{fofaeveq}, we still need three ingredients.

\begin{itemize}

\item Renormalization group invariance of the form factor requires
\beqa
 \left( \mu \, \frac{\partial}{\partial \mu}  + \beta (\e, \as) 
 \frac{\partial}{\partial \as} \right) G \left(\frac{Q^2}{\mu^2}, \as(\mu^2), 
 \e \right) & = & \nonumber \\ - \, \beta(\e, \as) \frac{\partial}{\partial \as} 
 K \left(\e, \as(\mu^2) \right) & \equiv & \gamma_K
 \left(\alpha_s (\mu^2) \right)~, 
 \label{defgamk}
\eeqa
where in the second equation we have used the fact that $K$ is a pure 
counterterm, and thus has no explicit scale dependence. \eq{defgamk} defines
the anomalous dimension $\gamma_K$, and allows one to solve for the 
$\mu$ dependence of $G$ in terms of an initial condition, say at $\mu = Q$.

\item The infrared freedom of the theory for $\e < 0$ provides us with a
simple initial condition for \eq{fofaeveq},
\beq
\overline{\alpha} (\mu^2 = 0, \e < 0) 
= 0~~ \rightarrow \Gamma \left( 0, \as(\mu^2), \e \right) = 
\Gamma \left( 1, \overline{\alpha}\left(0, \e \right), \e \right) = 1~.
\eeq

\item By the same token, the counterterm $K$ can be expressed directly as an integral of the anomalous dimension $\gamma_K$, using \eq{defgamk} and the vanishing of the coupling at  $\mu^2 = 0$. One verifies that
\beq
  K \left(\e, \as (\mu^2) \right) = - \frac{1}{2} \int_0^{\mu^2}
  \frac{d \lambda^2}{\lambda^2} \gamma_K \left(\bar{\alpha} 
  (\lambda^2, \e) \right)~.
\label{solevoK}
\eeq

\end{itemize}

\noindent Putting these ingredients together, one can express the solution to 
\eq{fofaeveq} in a simplified form, displaying the fact that infrared and 
collinear poles to all orders are generated by just two functions of the 
coupling, $G$ and $\gamma_K$. One finds~\cite{Dixon:2008gr}
\beqa
  \Gamma \left( Q^2, \e \right) & = & \exp \left\{ \frac{1}{2} 
  \int_0^{- Q^2} \frac{d \xi^2}{\xi^2} \left[
  G \Big(-1, \overline{\alpha}  \left(\xi^2, \e \right), \e \Big) \right. \right. 
  \nonumber \\
  && \left. \left. - \frac{1}{2} \, \gamma_K \Big( \overline{\alpha}
  \left(\xi^2, \e \right) \Big) \, \log \left(\frac{- Q^2}{\xi^2} \right) \right] 
  \right\} \, .
\label{sol2} 
\eeqa
In light of recent developments, both in QCD and in ${\cal N} = 4$ SYM, it is 
worth emphasizing that the form factors play an important role also in the 
much more general case of fixed-angle scattering amplitudes with any number 
of external legs, for massless gauge theories. Such amplitudes also factorize in 
a manner similar to \eq{factorize}, albeit with a more complicated color 
structure. Indeed, an amplitude with $m$ external colored legs, 
${\cal M}_{\{a_i\}}$, $i = 1, \ldots, m$,  can be written as a vector in the 
space of available color configuration,  with components ${\cal M}^{(m)}_L$ 
in a suitable basis of color tensors $c^L_{\{a_i\}}$. One may then  
write~\cite{Sterman:2002qn}
\beqa
\label{facamp}
  \M^{[m]}_{L} \left(\beta_j, \frac{Q^2}{\mu^2}, \as(\mu^2),
  \epsilon \right) \hspace{-1pt} & = & \hspace{-1pt} \prod_{i = 1}^{m} J_i
  \left(\frac{Q'{}^2}{\mu^2},\as(\mu^2),\epsilon \right)
  S^{[m]}_{L K} \left(\beta_j, \frac{Q'{}^2}{\mu^2}, \frac{Q'{}^2}{Q^2},
  \as(\mu^2), \epsilon \right) \nonumber \\
  &\ & \hspace{0.2mm} \times \ H^{[m]}_K \left(\beta_j,
  \frac{Q^2}{\mu^2}, \frac{Q'{}^2}{Q^2}, \as(\mu^2) \right) \, .
\eeqa
\eq{facamp} is expressed in terms of velocity four-vectors $\beta_i$ for 
each external leg, and the restriction to fixed-angle scattering has been 
exploited to extract from particle momenta a common hard scale $Q$; the 
scale $Q'$,  on the other hand, plays the role of a factorization scale 
separating infrared and collinear momenta; collinear singularities are 
organized into the $m$ `jet' functions $J_i$, each characterized only 
by the properties of the originating parton; soft gluons, on the other hand, 
can mix the color components of the hard scattering and thus are organized 
into a matrix $S^{[m]}_{L K}$, acting on a vector of finite coefficient  functions 
$H^{[m]}_K$. One may now exploit the fact that the jets $J_i$ collect the same 
collinear and infrared-collinear singular regions as the form factors $\Gamma_i$ 
for the same parton species: soft wide-angle radiation would be different, 
but one can make use of the fact that the soft matrix $S$ is defined up to
multiplication times a multiple of the identity matrix in order to reconstruct
the appropriate soft emission structure. In other words, there exists a 
factorization scheme such that one can define
\begin{eqnarray}
\label{jsudakov}
  J_i \left(\frac{Q'{}^2}{\mu^2}, \as(\mu^2), \epsilon\right)
  &=& \left[\Gamma_i
  \left(\frac{Q'{}^2}{\mu^2}, \as(\mu^2), \epsilon\right)
  \right]^{\frac{1}{2}} \, .
\end{eqnarray}
This factorization is especially useful because it teaches us that the structure
of collinear singularities of fixed-angle multi-leg scattering amplitudes is 
completely captured by partonic form factors. Furthermore, in this factorization scheme, the matrix $S$ becomes proportional to the identity matrix in the
planar, $N_c \to \infty$ limit. This feature simplifies considerably the analysis
in the interesting case of planar ${\cal N} = 4$ SYM, where a continuation of 
the amplitude to strong coupling has, in some cases, become possible.

\section{Beyond QCD: conformal gauge theories}
\label{confo}

One striking feature of \eq{sol2} is the fact that the logarithm of the form 
factor is expressed in terms of two {\it finite} functions of the coupling, 
$G$ and $\gamma_K$. All infrared and collinear poles are generated by 
the explicit integration over the scale of the running coupling. In QCD, 
and for $\e < 0$, the scale dependence of the coupling (see for example
\eq{highoas}) is such that poles up to $1/\e^{p+1}$ are generated at 
order $\as^p$. By contrast, in a conformal gauge theory such as 
${\cal N} = 4$ SYM, regularized by dimensional continuation, the 
coupling runs simply according to its engineering dimension in 
$d = 4 - 2 \e$; as a consequence, the integration in 
\eq{sol2} yields at most double poles. Expanding $\gamma_K$ and 
$G$ in powers of $\as/\pi$, and denoting their perturbative coefficients 
by $\gamma_K^{(n)}$ and $G^{(n)} (\e)$ respectively, one easily 
finds~\cite{Bern:2005iz}
\beqa
  \log \left[ \Gamma \left( \frac{Q^2}{\mu^2}, \as(\mu^2),  \e \right) 
  \right] & = & - \frac{1}{2} \sum_{n = 1}^\infty \left( \frac{\as 
  (\mu^2)}{\pi} \right)^n  \left( \frac{\mu^2}{- Q^2} \right)^{n \e} \left[ 
  \frac{\gamma_K^{(n)}}{2 n^2 \e^2} + \frac{G^{(n)} (\e)}{n \e} \right]
  \nonumber \\ & = & 
  - \frac{1}{2} \sum_{n = 1}^\infty \left( \frac{\as (Q^2) }{\pi} 
  \right)^n  {\rm e}^{- {\rm i} \pi n \e} \left[ 
  \frac{\gamma_K^{(n)}}{2 n^2 \e^2} + \frac{G^{(n)} (\e)}{n \e} \right]~,
\label{confgam}
\eeqa
where in the second line I note that the logarithm of the form factor 
displays exact renormalization group invariance, as expected. Using 
\eq{confgam}, it is possible to study the analytic continuation of 
the form factor from time-like to space-like kinematics, in the conformal 
case. This continuation is of practical interest in QCD: in fact, as shown to 
all orders in~\cite{Magnea:1990zb}, the modulus of the ratio of the 
time-like to the space-like form factor is finite in $d = 4$; furthermore, 
this ratio is closely related to physically observable cross sections: 
for example, it resums a class of large constant contributions to 
the Drell-Yan cross section in the DIS factorization 
scheme~\cite{Parisi:1979xd,Sterman:1986aj,Eynck:2003fn}. In the 
present case, having constructed a finite quantity, one may take $\e \to 0$
and compute the ratio in the four-dimensional theory with exact conformal
invariance. One finds~\cite{Dixon:2008gr}
\beq
  \left| \frac{\Gamma(Q^2)}{\Gamma(- Q^2)} \right|^2 = \exp \left[ 
  \frac{\pi^2}{4} \, \gamma_K \left( \alpha_s (Q^2) \right) \right]~.
\label{exact}
\eeq
\eq{exact} resums perturbation theory for finite quantities which admit 
a non-perturbative definition in terms of operator matrix elements. Thus,
it can be conjectured to be an exact result. It would be of great 
interest if a strong-coupling analogue could be derived.

Combining \eq{facamp}, \eq{jsudakov} and \eq{confgam} places strong constraints on the all-order structure of scattering amplitudes in 
dimensionaly-regularized ${\cal N} = 4$ SYM. In fact, after a reanalysis 
of one- and two-loop results for the four-point planar MHV 
amplitude~\cite{Anastasiou:2003kj}, Bern, Dixon and Smirnov (BDS)
performed the highly non-trivial calculation of the same amplitude at three 
loops~\cite{Bern:2005iz}, and found an intriguing pattern of exponentiation,
consistent with \eq{confgam}, but extending to non-singular, 
$\e$-independent terms. In order to illustrate this pattern, note that in the 
planar limit the all-order factorized matrix element in \eq{facamp} has color
structure proportional to the tree-level amplitude; in this case the soft matrix 
$S$ can be taken to be diagonal. Defining then, in shorthand notation, a 
reduced matrix element $\widetilde{\cal M}^{[m]} (\e)$, by dividing out the 
tree-level result, one can show that the non-abelian exponentiation following 
from \eq{facamp} and \eq{jsudakov} leads to the 
expression~\cite{Bern:2005iz}
\beq
\hspace{-8mm}
  \widetilde{\cal M}^{[m]} (\e) = \exp \left[ \sum_{p = 1}^\infty
  \left(\frac{\lambda}{8 \pi^2} \right)^p \left( f^{(p)} (\e) \,  
  M_1^{[m]} (p \, \e) + h^{[m]}_p (k_i)  + {\cal O} (\e) 
  \right) \right] \, .
\label{fromlance}
\eeq
Here $\lambda$ is the 't Hooft coupling, $\lambda = g^2 N_c$; 
$M_1^{[m]}$ is the one-loop amplitude, which however is evaluated with a
rescaled value of $\e$ (a feature clearly visible in \eq{confgam}); 
$f^{(p)} (\e)$ is a quadratic polynomial in $\e$, with constant and linear 
terms determined by \eq{confgam}, 
\beq
f^{(p)} (\e) = \sum_{n = 0}^2 f^{(p)}_n \e^n \, , \qquad
f^{(p)}_0 = \frac{\gamma_K^{(p)}}{4} \, , \quad 
f^{(p)}_1 = \frac{p}{2} \, G^{(p)} (0) \, ,
\label{fdix}
\eeq
while $f_2^{(p)}$ can be determined by consistency, considering the case 
in which subsets of external momenta become collinear; finally, 
$h^{[m]}_p (k_i)$ is a finite remainder, which in a general gauge 
theory depends both on the number of particles $m$ and on their momenta
$k_i$. BDS~\cite{Bern:2005iz} observed that the finite remainder 
$h^{(4)}_p (k_i)$ is a constant, independent of kinematics, for
$p \leq 3$; using also results on collinear limits derived 
in~\cite{Anastasiou:2003kj}, 
they conjectured that this property might remain true to 
all orders in the 't Hooft coupling and for any number $m$ of particles. On
the other hand, Alday and Maldacena~\cite{Alday:2007hr}, computing the 
four-point amplitude at strong 't Hooft coupling by means of the AdS-CFT
correspondence, found a structure closely matching \eq{fromlance}.
These results have lead to a sustained effort by several groups, employing
rather different theoretical tools, to study the structure of amplitudes in 
${\cal N} = 4$ SYM and related theories, and to constrain and compute the anomalous dimensions $\gamma_K$ and $G$ that govern their singularities 
(for references, see the reviews in~\cite{Dixon:2008tu,Alday:2008cg}).
To summarize very briefly the status of these efforts to date, the 
BDS conjecture is now expected to hold for the four- and -five point 
amplitudes, while it is known to break down for the six-point 
amplitude, starting at two loops~\cite{Bern:2008ap,Drummond:2008aq}; an 
ansatz exists~\cite{Beisert:2006ez} for the function $\gamma_K (\lambda)$, 
which reproduces all available perturbative results, both at weak and at 
strong coupling~\cite{Benna:2006nd,Basso:2007wd}; the function $G$ has 
also been analyzed in detail~\cite{Dixon:2008gr}, in the general case of an
arbitrary massless gauge theory, expressing it in terms of anomalous 
dimensions of operators involving Wilson lines and fundamental fields, plus
running coupling contributions. For a conformal gauge theory, one finds 
the very simple result
\beq
  G (1, \as, \e = 0) = 
  2 B_\delta \left( \as \right)  + G_{\rm eik} \left( \as \right)~,
\label{finG}
\eeq
where $G_{\rm eik}$ is a subleading anomalous dimension associated
with Wilson lines, and thus in principle amenable, like $\gamma_K$, to studies 
with non-perturbative techniques, while $B_\delta$ is the virtual contribution 
to the Altarelli-Parisi splitting kernel. $B_\delta$ involves matrix elements
of local fields as well as Wilson lines, so it would be quite interesting to
to see how an equation of the form of \eq{finG} might arise at strong 
coupling.

\section{Conclusion}

The study of long-distance singularities of gauge theories began more 
than seventy years ago~\cite{Bloch:1937pw}, yet it remains an active 
and fertile field of research. In QCD, all-order results for soft and collinear 
gluons are instrumental for phenomenology, providing nontrivial tests of 
finite order calculations, and forming the basis for the resummation of
several classes of large logarithms that would otherwise hinder the 
applicability of perturbation theory. From a theoretical standpoint, studying
long-distance effects to all orders in perturbation theory opens a window on
non-perturbative effects, which are suppressed by powers of the hard scale
but may still be very relevant for high-energy cross sections in certain 
kinematical regimes. When tools are available for a quantitative study of
a gauge theory at strong coupling, as is the case for maximally 
supersymmetric Yang-Mills theory, soft and collinear singularities of 
amplitudes still provide a bridge between weak coupling and non-perturbative
regimes. Recent progress in the study of ${\cal N} = 4$ SYM has been 
especially remarkable; bringing together tools from perturbation theory,
string theory and integrable models, it has been possible to reach results
that point towards a very ambitious goal: a full and detailed understanding 
of a non-trivial four-dimensional gauge theory. We may indeed look forward 
to new developments and applications, both on the practical side of collider
phenomenology, and on our way to a deeper understanding of quantum 
field theory.


\vspace{-1mm}
\begin{thebibliography}{99}

\bibitem{Kinoshita:1962ur}
  T.~Kinoshita,
  %``Mass Singularities Of Feynman Amplitudes,''
  J.\ Math.\ Phys.\  {\bf 3} (1962) 650.
  %%CITATION = JMAPA,3,650;%%

\bibitem{Lee:1964is}
  T.~D.~Lee and M.~Nauenberg,
  %``Degenerate Systems and Mass Singularities,''
  Phys.\ Rev.\  {\bf 133} (1964) B1549.
  %%CITATION = PHRVA,133,B1549;%%

\bibitem{Collins:1989gx}
  J.~C.~Collins, D.~E.~Soper and G.~Sterman,
  %``Factorization of Hard Processes in QCD,''
  Adv.\ Ser.\ Direct.\ High Energy Phys.\  {\bf 5} (1988) 1,
  hep-ph/0409313.
  %%CITATION = 00319,5,1;%%

\bibitem{Altarelli:1977zs}
  G.~Altarelli and G.~Parisi,
  %``Asymptotic Freedom In Parton Language,''
  Nucl.\ Phys.\  B {\bf 126} (1977) 298.
  %%CITATION = NUPHA,B126,298;%%

\bibitem{Contopanagos:1996nh}
  H.~Contopanagos, E.~Laenen and G.~Sterman,
  %``Sudakov factorization and resummation,''
  Nucl.\ Phys.\  B {\bf 484} (1997) 303,
  hep-ph/9604313.
  %%CITATION = NUPHA,B484,303;%%

\bibitem{Laenen:2004pm}
  E.~Laenen,
  %``Resummation for observables at TeV colliders,''
  Pramana {\bf 63} (2004) 1225.
  %%CITATION = PRAMC,63,1225;%%

\bibitem{Becher:2006nr}
  T.~Becher and M.~Neubert,
  %``Threshold resummation in momentum space from effective field theory,''
  Phys.\ Rev.\ Lett.\  {\bf 97} (2006) 082001, hep-ph/0605050.
  %%CITATION = PRLTA,97,082001;%%

\bibitem{Dasgupta:2003iq}
  M.~Dasgupta and G.~P.~Salam,
  %``Event shapes in e+ e- annihilation and deep inelastic scattering,''
  J.\ Phys.\ G {\bf 30} (2004) R143, hep-ph/0312283.
  %%CITATION = JPHGB,G30,R143;%%

\bibitem{Gardi:2006jc}
  E.~Gardi,
  %``Inclusive distributions near kinematic thresholds,''
  %in {\it Proceedings of FRIF workshop on first principles non-perturbative 
  %QCD of hadron jets, LPTHE, Paris, France, 12-14 Jan 2006, pp E003}, 
  hep-ph/0606080.
  %%CITATION = ECONF,C0601121,E003;%%

\bibitem{Alday:2007hr}
  L.~F.~Alday and J.~M.~Maldacena,
  %``Gluon scattering amplitudes at strong coupling,''
  JHEP {\bf 0706} (2007) 064, arXiv:0705.0303 [hep-th].
  %%CITATION = JHEPA,0706,064;%%

\bibitem{Dixon:2008tu}
  L.~J.~Dixon,
  %``Gluon scattering in N=4 super-Yang-Mills theory from weak to strong
  %coupling,''
  arXiv:0803.2475 [hep-th].
  %%CITATION = ARXIV:0803.2475;%%

\bibitem{Alday:2008cg}
  L.~F.~Alday,
  %``Lectures on Scattering Amplitudes via AdS/CFT,''
  arXiv:0804.0951 [hep-th].
  %%CITATION = ARXIV:0804.0951;%%

\bibitem{Dixon:2008gr}
  L.~J.~Dixon, L.~Magnea and G.~Sterman,
  %``Universal structure of subleading infrared poles in gauge theory
  %amplitudes,''
  arXiv:0805.3515 [hep-ph].
  %%CITATION = ARXIV:0805.3515;%%

\bibitem{Forte:2002ni}
  S.~Forte and G.~Ridolfi,
  %``Renormalization group approach to soft gluon resummation,''
  Nucl.\ Phys.\  B {\bf 650} (2003) 229, hep-ph/0209154.
  %%CITATION = NUPHA,B650,229;%%

\bibitem{Collins:1989bt}
  J.~C.~Collins,
  %``Sudakov form factors,''
  Adv.\ Ser.\ Direct.\ High Energy Phys.\  {\bf 5} (1989) 573,
  hep-ph/0312336.
  %%CITATION = 00319,5,573;%%

\bibitem{Magnea:1990zb}
  L.~Magnea and G.~Sterman,
  %``Analytic continuation of the Sudakov form-factor in QCD,''
  Phys.\ Rev.\  D {\bf 42} (1990) 4222.
  %%CITATION = PHRVA,D42,4222;%%

\bibitem{Magnea:2000ss}
  L.~Magnea,
  %``Analytic resummation for the quark form factor in QCD,''
  Nucl.\ Phys.\  B {\bf 593} (2001) 269, hep-ph/0006255.
  %%CITATION = NUPHA,B593,269;%%

\bibitem{Sterman:2002qn}
  G.~Sterman and M.~E.~Tejeda-Yeomans,
  %``Multi-loop amplitudes and resummation,''
  Phys.\ Lett.\  B {\bf 552} (2003) 48, hep-ph/0210130.
  %%CITATION = PHLTA,B552,48;%%

\bibitem{Bern:2005iz}
  Z.~Bern, L.~J.~Dixon and V.~A.~Smirnov,
  %``Iteration of planar amplitudes in maximally supersymmetric Yang-Mills
  %theory at three loops and beyond,''
  Phys.\ Rev.\  D {\bf 72} (2005) 085001, hep-th/0505205.
  %%CITATION = PHRVA,D72,085001;%%

\bibitem{Parisi:1979xd}
  G.~Parisi,
  %``Summing Large Perturbative Corrections In QCD,''
  Phys.\ Lett.\  B {\bf 90} (1980) 295.
  %%CITATION = PHLTA,B90,295;%%

\bibitem{Sterman:1986aj}
  G.~Sterman,
  %``Summation of Large Corrections to Short Distance Hadronic Cross-Sections,''
  Nucl.\ Phys.\  B {\bf 281} (1987) 310.
  %%CITATION = NUPHA,B281,310;%%

\bibitem{Eynck:2003fn}
  T.~O.~Eynck, E.~Laenen and L.~Magnea,
  %``Exponentiation of the Drell-Yan cross section near partonic threshold  in
  %the DIS and MS-bar schemes,''
  JHEP {\bf 0306} (2003) 057, hep-ph/0305179.
  %%CITATION = JHEPA,0306,057;%%

\bibitem{Anastasiou:2003kj}
  C.~Anastasiou, Z.~Bern, L.~J.~Dixon and D.~A.~Kosower,
  %``Planar amplitudes in maximally supersymmetric Yang-Mills theory,''
  Phys.\ Rev.\ Lett.\  {\bf 91} (2003) 251602, hep-th/0309040.
  %%CITATION = PRLTA,91,251602;%%

\bibitem{Bern:2008ap}
  Z.~Bern {\it et al.},
  %``The Two-Loop Six-Gluon MHV Amplitude in Maximally Supersymmetric 
  %Yang-Mills Theory,''
  arXiv:0803.1465 [hep-th].
  %%CITATION = ARXIV:0803.1465;%%

\bibitem{Drummond:2008aq}
  J.~M.~Drummond, J.~Henn, G.~P.~Korchemsky and E.~Sokatchev,
  %``Hexagon Wilson loop = six-gluon MHV amplitude,''
  arXiv:0803.1466 [hep-th].
  %%CITATION = ARXIV:0803.1466;%%

\bibitem{Beisert:2006ez}
  N.~Beisert, B.~Eden and M.~Staudacher, J. Stat. Mech. {\bf 0701} (2007)
  P02, hep-th/0610251.
  %%CITATION = JSTAT,0701,P021;%%

\bibitem{Benna:2006nd}
  M.~K.~Benna, S.~Benvenuti, I.~R.~Klebanov and A.~Scardicchio,
  {\it Phys. Rev. Lett.} {\bf 98} (2007) 131603, hep-th/0611135.
  %%CITATION = PRLTA,98,131603;%%

\bibitem{Basso:2007wd} 
  B.~Basso, G.~P.~Korchemsky and J.~Kota\'nski,
  {\it Phys. Rev. Lett.} {\bf 100} (2008) 091601,
  arXiv:0708.3933 [hep-th].
  %%CITATION = ARXIV:0708.3933;%%

\bibitem{Bloch:1937pw}
  F.~Bloch and A.~Nordsieck,
  %``Note on the Radiation Field of the electron,''
  Phys.\ Rev.\  {\bf 52} (1937) 54.
  %%CITATION = PHRVA,52,54;%%

\end{thebibliography}
\end{document}